\documentclass{elsart}

\usepackage{natbib}

 \usepackage{epsfig}

\usepackage{amssymb}
\usepackage{graphicx}

\begin{document}

\begin{frontmatter}


%\title{Title\thanksref{label1}}

\title{EFFECTIVENESS OF MEASURES OF PERFORMANCE DURING SPECULATIVE
BUBBLES}
\author{Filippo Petroni}
\address{GRAPES, B5, Sart-Tilman, B-4000 Liege, Belgium}
\ead{fpetroni@gmail.com}
\author{Giulia Rotundo}
\address{Faculty of Economics, University of Tuscia, Viterbo}
\address{Faculty of Economics, University of Rome ``La Sapienza''}
\ead{giulia.rotundo@uniroma1.it}

\begin{abstract}
Statistical analysis of financial data most focused on testing
the validity of Brownian motion (Bm). Analysis performed on
several time series have shown deviation from the Bm hypothesis,
that is at the base of the evaluation of many financial
derivatives. We inquiry in the behavior of measures of performance
based on maximum drawdown movements (MDD), testing their stability
when the underlying process deviates from the Bm hypothesis. In
particular we consider the fractional Brownian motion (fBm), and
fluctuations estimated empirically on raw market data. The case
study of the rising part of speculative bubbles is reported.
\end{abstract}


\end{frontmatter}

 \section{Introduction}
Hypothesis testing in time series modeling is a delicate step
that relies on the precision of statistical analysis on raw data.
\newline
This is most true in the field of  empirical finance, where
information spreading and investors' behavior are rendered into
statistical properties that market data should verify. In this
framework, the analysis of correlation in raw market data
constitutes a base step for model identification. The slow decay
of autocorrelation has been evidenced through many empirical
studies, and new methods were developed to measure it through the
Hurst ($H$) exponent \cite{DFA1,bj,ms}.  The outcome of
statistical analysis on data often shows oscillations of $H$
outside the confidence intervals \cite{weron}, that seem to
exclude the Bm hypothesis, and to assess the compatibility of data
with fBm. It is worth remarking that the values of $H$ are highly
sensitive to the time window selected, internal time windows
showing the widest variety of fluctuations. Therefore, also the
fBm hypothesis constitutes a first raw approximation for modeling
the entire data sets available. Moreover, even without numerical
problems on the estimation of $H$, the slow correlation decay is
compatible with the most classical fractional Brownian motion as
well as with other processes having an autocorrelation function
depending on the time lag \cite{em,BS_FPMAGR}. Recently, it has
been shown theoretically that the same kind of correlation decay
can rise also from Markov models \cite{JMCHurst,jmc2} as well as
from models of relaxation with a slowly vanishing correlation
\cite{lapas}. To distinguish between Markovian vs non-Markovian
behavior is relevant for its implication in term of market risk,
and subsequent approach to portfolio management and performance
measures. \newline The occurrence of either Bm or geometric
Brownian motion (gBm) opens the way to many theoretical results,
often expressed through fast to compute closed form formulas.
Therefore, it is worth using Bm and/or gBm as a benchmarks, and to
understand the behaviour of derivatives based on them. \newline We
aim at exploring the limits of validity of the theoretical
estimate of performance measures derived in the Bm case. We focus
on measures of performance based on maximum drawdown. More in
details, the first task is to inquiry through simulations the
distance of measures of performance on fBm data from the
theoretical estimate of measures of performance that are available
for Bm.  The second task is to examine the case study of the
rising part of speculative bubbles. The first step here is to
refine the analysis on the Gaussianity of increments varying the
time window, so to examine more in details a necessary condition
for the fBm modeling. The second step is to examine the risk
measures and the related performance indices during the rising
part of the bubbles, and to compare them with the theoretical
estimates that hold for (f)Bm.
\newline The paper is organized as follows. The next section
resumes the risk measures, and performance indices based on
maximum drawdown. Section 3. shows the results of the simulations
performed on the fBm. Section 4. sums up the speculative bubble
approach,  it shows the behavior of risk and performance  measures
on the rising part of speculative bubbles, and a comparison of them
with the theoretical Bm results is also performed.

\section{Performance indices based on the maximum drawdown}
The analysis of long sequences of drawdown market movements is
relevant in itself, because long lasting decreasing trends can
force small investors to abandon the market, and they can cause
fund managers to loose clients. One of the most widely quoted
measures of risk used by hedge funds and commodity traders
advisors is peak to trough drawdown \cite{bdl}, i.e. the size of
the largest loss.
 \newline Using a stochastic processes approach, the maximum drawdown at time $T$ of a random process on $[0,T]$ can be defined
 formally as follows.
\newline
 {\bf Definition 1} Let $\{X(t)\}$ be a stochastic process. The maximum drawdown $MDD$ is given by
\begin{equation} \label{mdd}
MDD(T)=sup_{t \in [0,T]} ( sup_{s \in [0,t]} X(s) - X(t)).
\end{equation}
Ismail et al. (2005) comprehensively analyzed maximum drawdown as
a measure of risk with emphasis on updated ratios and Brownian
motion \cite{iarisk,mia}.  Assume that the value of a portfolio
follows a gBm:
\begin{equation} \label{gBm}
ds=\hat{\mu} s dt + \hat{\sigma} s dW, \,\, 0\le t \le T
\end{equation}
Taking a logarithmic transformation $x=log s$, then $x$ follows a Bm.
Defining $\mu=\hat{\mu} - \frac{1}{2} \hat{\sigma}^2$, and $
\sigma=\hat{\sigma}$:
\[dx=\mu dt + \sigma dW, \,\, 0\le t \le T\]
where time is measured in years, $\mu$ is the average return
per unit time, $\sigma$ is the standard deviation of the returns
per unit time and $dW$ is the usual Wiener increment.
\newline
Theoretical estimates of the expected value of $MDD(T)$, $E(MDD)$,
have been shown in the Bm case \cite{iarisk,mia}:
\begin{equation} \label{EMDD}
E(MDD(T))=\left\{
\begin{array}{lr}
\frac{2\sigma^2}{\mu} Q_p(\frac{2\mu^2T}{2\sigma^2}
)\rightarrow_{T\rightarrow \infty} \frac{2\sigma^2}{\mu}\left(
0.63+0.5logT+log\frac{\mu}{\sigma}\right) & if \mu>0 \\ 1.25
\sigma \sqrt{T}& if \mu=0 \\ \frac{2\sigma^2}{\mu}
Q_n(\frac{2\mu^2T}{2\sigma^2} )\rightarrow_{T\rightarrow \infty}
-\mu -  \frac{\sigma^2}{\mu} & if \mu<0 \\
\end{array}
\right.
\end{equation}
The need to consider the order of data alongside their most
classical statistical properties has already led to the
introduction of MDD into risk measures  (Drawdown at Risk (Dar)
and Conditional Drawdown at Risk (CDaR) alongside the most classic
VaR, ES and CVaR , A. Chekhlov et al. (2000))\cite{brandi,miao}.
\newline
The same need to accomplish the investors risk perception in
measures of performance has led to the modification of risk
adjustment in measures of performance of portfolios.
The classic Sharpe ratio is defined as follows:
\newline {\bf Definition 2} The Sharpe ratio is defined as
\begin{equation} Shrp(T)=\frac{\mu \,\, over \,\, [0,T]}{\sigma \,\, over \,\, [0,T]}
\end{equation}
The risk is measured through the standard deviation, and the
returns are adjusted w.r.t. it. Therefore, the Sharpe ratio is not
suitable for describing the adversion of investors to the
occurrence of downward movements. The MDD has been introduced
giving rise to measures of performance like the Calmar
ratio, that is defined as follows:
\newline
{\bf Definition 3} The Calmar ratio is a risk-adjusted measure of performance that is given by the formula:
\begin{equation}
Calmar(T)=\frac{\mu T}{E(MDD)}
\end{equation}
The Calmar ratio has been shown to be closely related to the
Sharpe ratio
\[Calmar(T)=\frac{\frac{T}{2} Shrp^2}{Q_p(\frac{T}{2}
Shrp^2)}\] The result \ref{EMDD} shows how the expected MDD is
related to the mean return and the standard deviation of the
returns in the Bm case, and it can be used at first to state that
\[\frac{E(MDD)}{\sigma}=\frac{2Q_p(\frac{T}{2}Shrp^2)}{Shrp}\]and then to conclude
that \cite{iarisk,mia}
\[Calmar(T)\rightarrow_{T\rightarrow \infty}  \frac{T
Shrp^2}{0.63+0.5logT+log(Shrp)}.
\]
Theoretical estimates are
not available for fBm or geometrical fBm (gfBm), and there is a
lack of sensitivity analysis to deviations from Bm and gBm
hypotheses.

\section{Measuring performance in the case of fBm}
 We aim at extending the results of
\cite{iarisk} to the case in which the underlying asset obeys a
fBm. %:
%\begin{equation} \label{gfBm}
%dX=\hat{\mu} dt + \hat{\sigma} dB_H, \,\, 0\le t \le T.
%\end{equation}
Some remarks are worth to be added about the suitability of
the measures listed in the previous section as a good tool for the
estimate of the performance when the underlying portfolio obeys a
fBm. The common characteristic of performance measures is to
consider return/risk ratios. Such ratios are different from each other because of the way in which the risk is defined. A relevant
remark is that the Sharpe ratio captures only the risk that can be
expressed through the variance.
\newline
We are going to consider the fBm, that is a Gaussian process. The main difference between fBm and
regular Brownian motion is that while the increments in Brownian
motion are independent, in fBm case they are dependent. However,
in both cases increments are Gaussian. It is well known that such
distribution is completely defined if the first and second order
moments are fixed. Therefore the Sharpe
 ratio totally captures the relevant moments of the probability distribution
 of increments also in the fBm case, and the differences from the Bm
 case are due only to the correlation structure. This statement does not hold for
other distributions of the increments and for other processes that
have been considered in literature for modeling the dynamics of
the returns. Weak stationarity is a necessary condition for the
validity of the above performance measures.  As an example, the
definition of Sharpe ratio does not represent entirely the risk
associated to the Martingale stochastic process examined in
\cite{jmc2}  due to the nonstationarity of the increments.
\newline
On the other hand, the Calmar ratio is not based explicitly on
the Gaussian assumption, that is needed only when the relationship
with the Sharpe ratio is examined. The estimate of the
relationship between the Calmar and the Sharpe ratio is mostly based
on the behavior of $E(MDD)$, that in (\ref{EMDD}) is obtained
through a theoretical approach, and through the estimate of the
function $Q(x)$. The main result is the detection of a phase
transition when $\mu$ changes its sign, and so $\mu=0$ is a
critical value. We start examining the behavior of $E(MDD)$ of fBm
through simulations.

\subsection{$E(MDD)$ in the fBm case}
\begin{figure}
(a)\includegraphics[height=6cm,width=6cm]{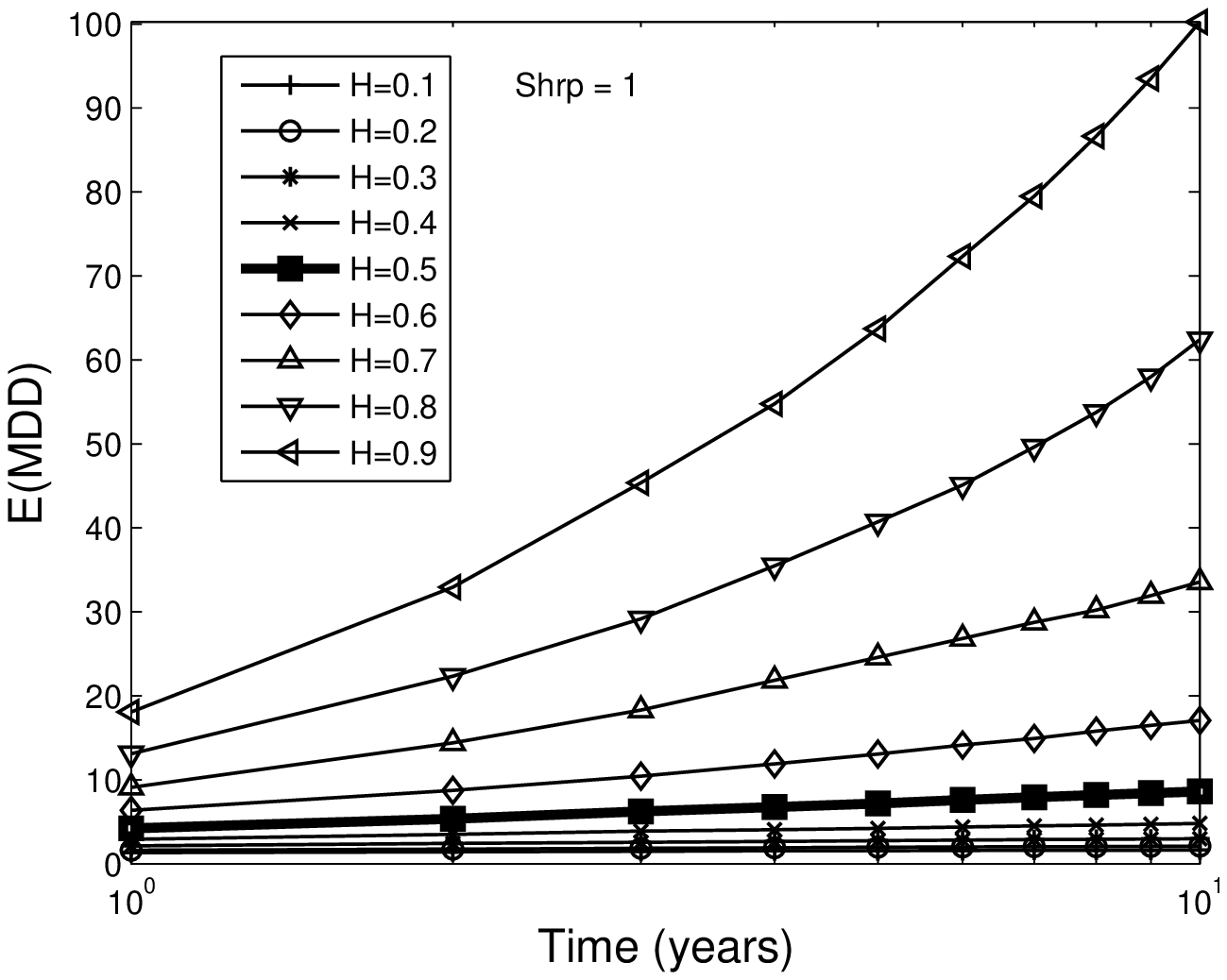}
(b)\includegraphics[height=6cm,width=6cm]{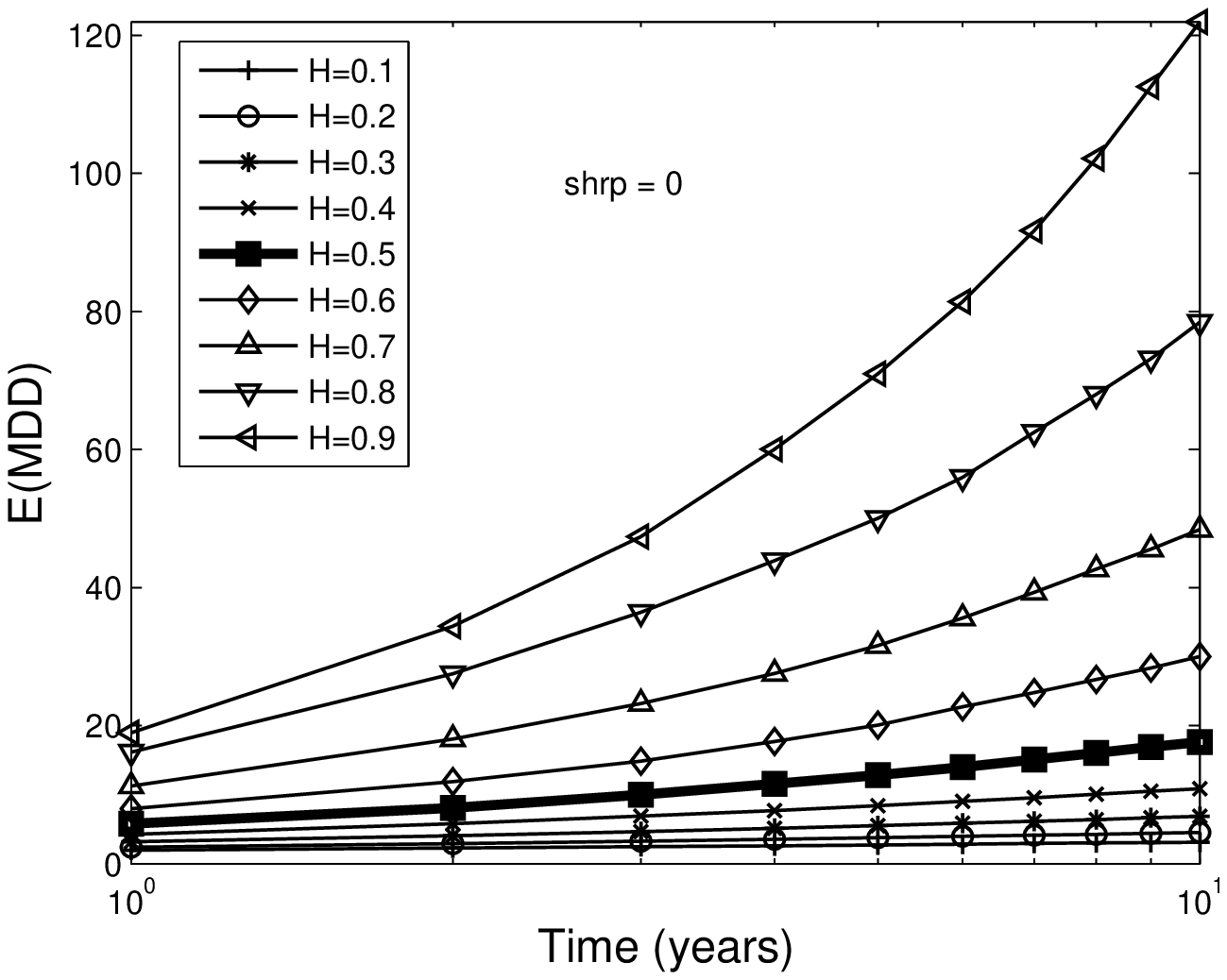}
(c)\includegraphics[height=6cm,width=6cm]{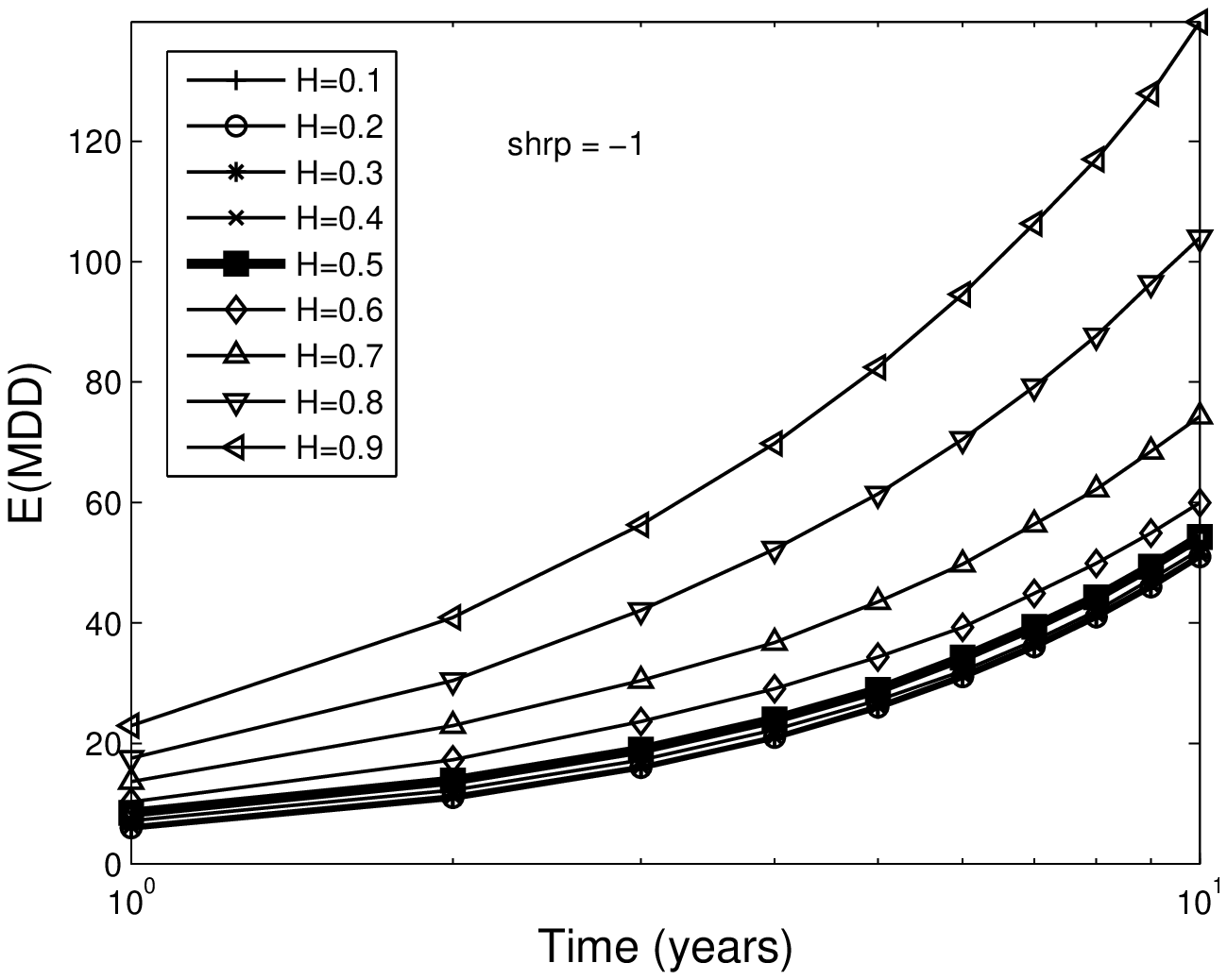}
(d)\includegraphics[height=3cm,width=6cm]{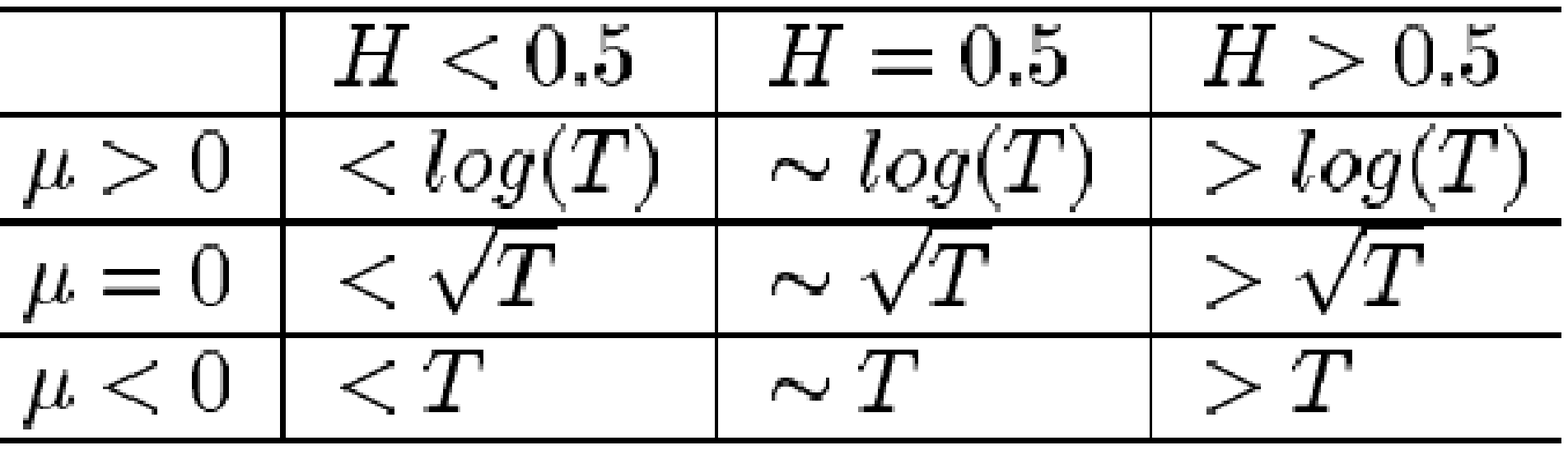}
%\begin{tabular}{l|l|l|l} \hline %\noalign{\smallskip}
% & $H<0.5$ & $H=0.5$ & $H>0.5$ \\ %\noalign{\smallskip}
% \hline
%$\mu>0$ & $<log(T)$ & $\sim log(T)$  &  $>log(T)$
%\\ %\noalign{\smallskip}
%\hline $\mu=0$ & $<\sqrt{T}$ & $\sim \sqrt{T}$  &  $>\sqrt{T}$
%\\ %\noalign{\smallskip}
%\hline $\mu<0$ & $ < T $ & $\sim T$  &  $ > T$
%\\ %\noalign{\smallskip}\hline
%\hline
%\end{tabular}

\caption{$x$-axes report the time expressed in years in a log
scale.  The Monte Carlo time for the synthetic time series is
measured in days and we considered one year as given by 256 days (see text).
For each $H \in \{0.1,0.2,0.3,0.4,0.5,0.6,0.7,0.8,0.9\}$, the
curves report the values of $E(MDD)$ for: {\bf (a)}
$\mu=\sigma=5$, corresponding to a return of $5\%$/year
considering a standard deviation of $5\%$/year ($Shrp=1$); {\bf
(b)} $\mu=0$, $\sigma=5$ ($Shrp=0$); {\bf (c)} $\mu=-5$,
$\sigma=5$ ($Shrp=-1$).
\newline {\bf (d)} In all three cases the behavior of $E(MDD)$ for $H=0.5$ separates the
regime for $H>0.5$ from the one for $H<0.5$.} \label{fig1}
\end{figure}

Simulations have been performed in order to estimate $E(MDD)$ on
fBm $X(t)$ by varying the Hurst exponent $H \in
\{0.1,0.2,0.3,0.4,0.5,0.6,0.7,0.8,0.9\}$. Figure \ref{fig1} shows the
behavior of $E(MDD)$ for the whole set of values of $H$
considered here. For each value of $H$ the figure reports the mean of
$E(MDD)$ calculated over 1000 fBm synthetic time series.
\newline
The Monte Carlo time for the synthetic time series is chosen such to be equivalent to days in real time series, we considered one year as given by 256 days (or Monte Carlo time steps). The plot is shown for a return of
$5\%$/year considering a standard deviation of $5\%$/year ($Shrp =
1$). Different values for $\mu$ and $\sigma$ in the range of real
market portfolios, i.e. the annual return $\mu\in [1,10]\%$ and
the annual volatility $\sigma \in [1,20] \%$ also give the same
behavior. Such range for $\mu$ gives rise to $Shrp>0$.
\newline
As expected the $H=0.5$ case is linear in the log-linear
plot of Figure \ref{fig1}.a. Some departure from linearity starts
already for $H=0.6$ and the discrepancy increases with increasing
$H$, for $H<0.5$ the curves are very close to each other and
to the linear one.
\newline
% Fig.  \ref{fig5} shows the distance ($\mid
%E(MDD(T))_{H}-E(MDD(T))_{0.5}\mid$) of each curve from the one
%drawn for $H=0.5$.
Therefore, when $\mu>0$, the curve of $E(MDD)$ for $H=0.5$
separates two different regimes; although we do not test any hypotheses on
the functional form of the curves in the fBm case, there is a clear passage from
curves slower than the logarithm to curves faster than the
logarithm.
\newline
This behavior is in good agreement with the persistency of fBm for
$H>0.5$ and for the mean reverting behavior for $H<0.5$. In the
former case long range drawdowns as well as drawups have a higher
probability to happen than in the latter case.
\newline
Figure  \ref{fig1}.b shows the case for $Shrp=0$. Also in this
case $H=0.5$ separates $E(MDD)$ growing faster ($H>0.5$) and
slower ($H<0.5$). Figure \ref{fig1}.c completes the analysis in the
case of $Shrp<0$, showing the same monotonic behavior of the
curves w.r.t. $H$. The complete scheme is summed up in Figure
\ref{fig1}.d.

\subsection{Performance measures in the fBm case}
\begin{figure}
\includegraphics[height=10cm,width=12cm]{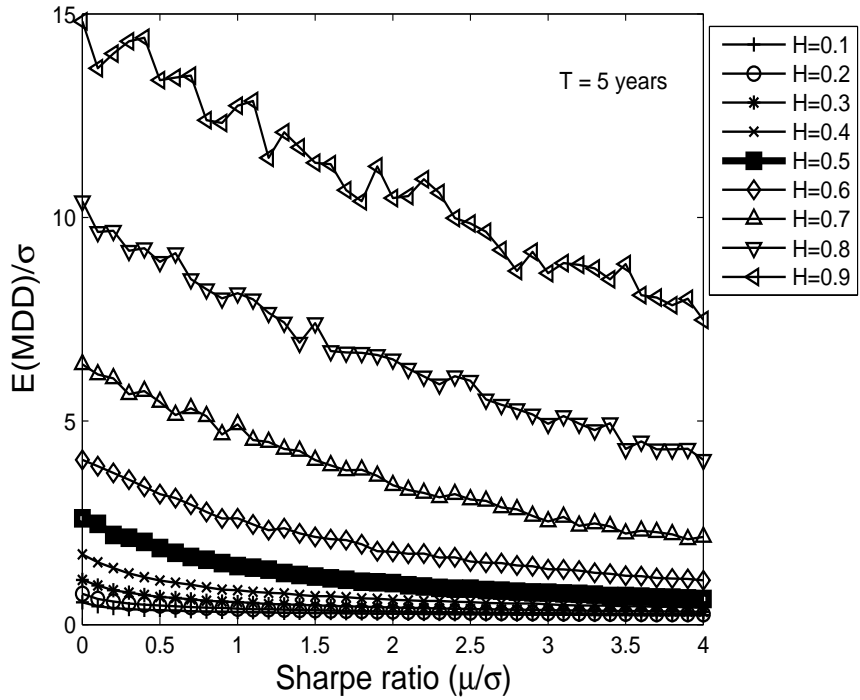}
\caption{The $E(MDD)$ per unit $\sigma$ versus Sharpe ratio
($\mu/\sigma$) for $\mu>0$. The time is fixed st $T=5 years$, and the
curves report the values for $H \in
\{0.1,0.2,0.3,0.4,0.5,0.6,0.7,0.8,0.9\}$. } \label{EMDDSharpe}
\end{figure}

\begin{figure}
(a)\includegraphics[height=6cm,width=6cm]{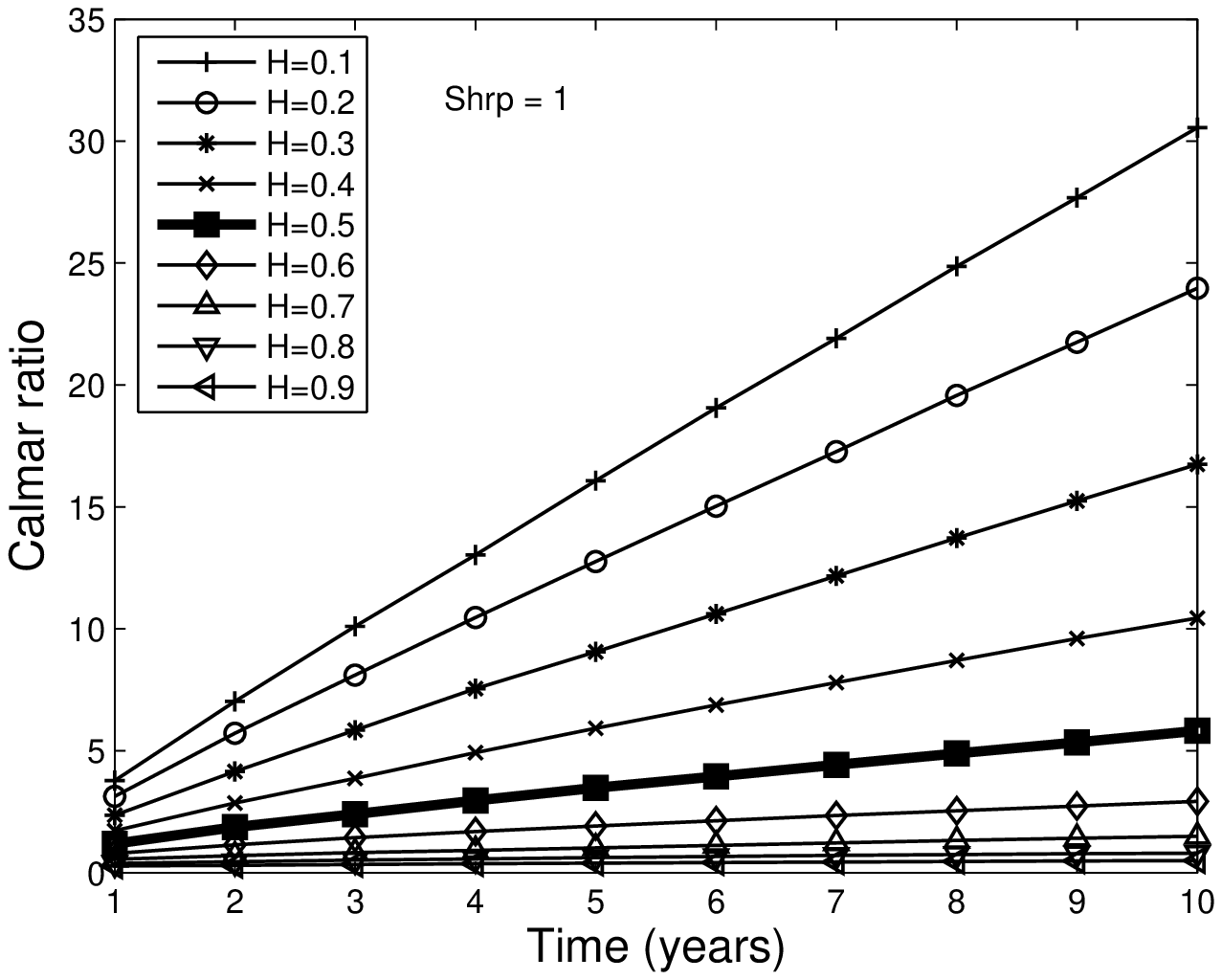}
(b)\includegraphics[height=6cm,width=6cm]{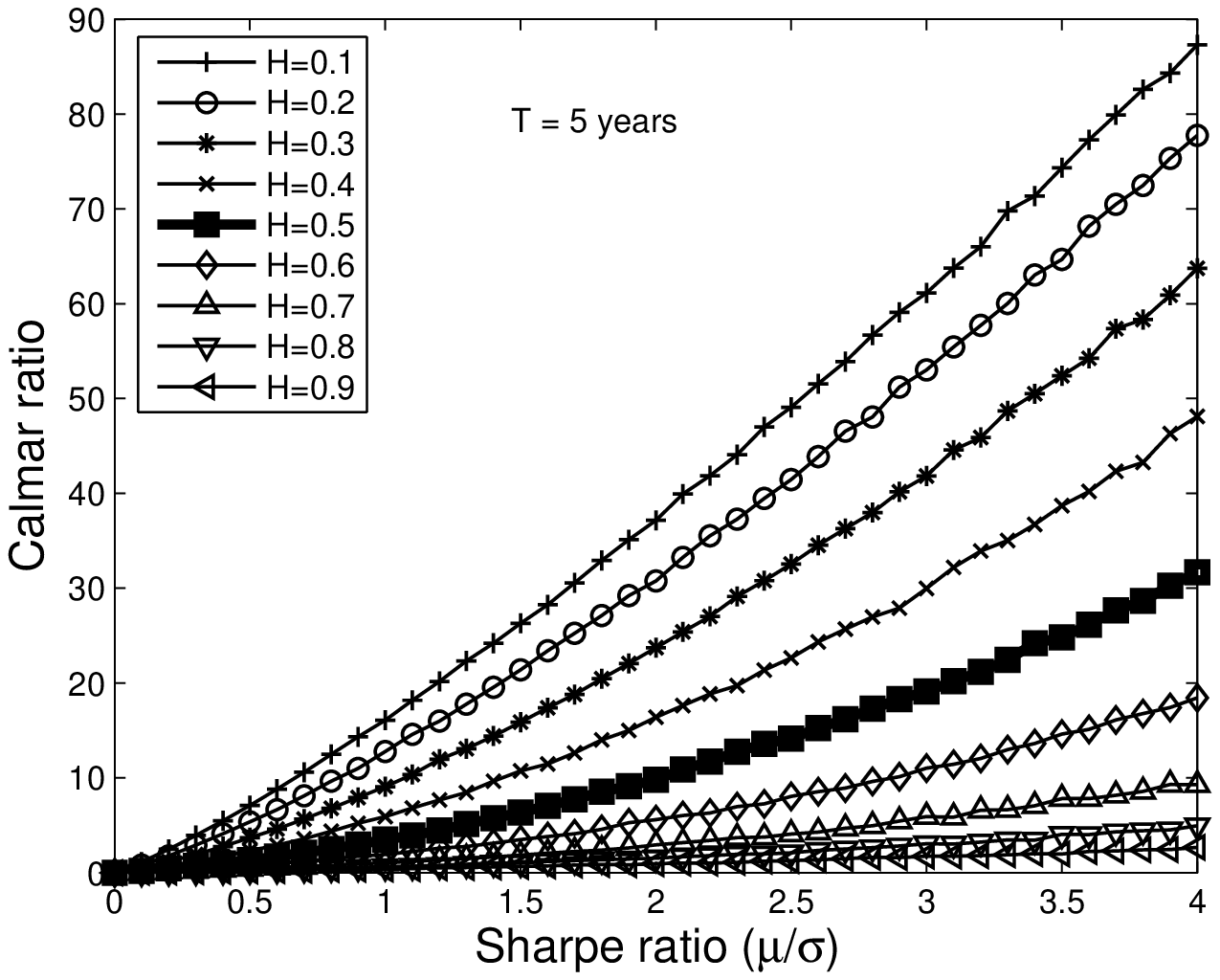} \caption{The
Calmar ratio (a) w.r.t. time; (b) as a function of the Sharpe
ratio for $\mu>0$. The case $H=0.5$ always separates the behavior
in case $H>0.5$ from $H<0.5$.} \label{fig3}
\end{figure}

Equation (\ref{EMDD}) relating $E(MDD)$ to $\frac{\mu}{\sigma}$
plays a key role in the assessment of the relationship between the
Calmar and the Sharpe ratios. Equation (\ref{EMDD}) was obtained
through a theoretical approach, supported by the numerical
estimate of the function $Q(x)$. In the case of fBm the behavior
of $E(MDD)$ w.r.t. time changes as shown in the previous
section. Therefore, the relationships between the Calmar and the
Sharpe ratios must be analyzed considering the new behavior of
$E(MDD)$. From now on, we focus on the more interesting case of
profitable portfolios, i.e. $\mu>0$.
\newline
Figure \ref{EMDDSharpe} reports the behavior of the Sharpe ratio as a function
of $E(MDD)$ for each $H \in
\{0.1,0.2,0.3,0.4,0.5,0.6,0.7,0.8,0.9\}$. Figure \ref{fig3}
considers the Calmar ratio, and it shows its behavior by varying
H. In both cases $H=0.5$ separates the two regimes $H>0.5$ from
$H<0.5$ and the behavior is monotonic w.r.t. $H$.
%next lavoro:
%Such analyses allow to estimate the new relationship for the time
%conversion factor STIMARE IL COEFF DI PROP CALMAR barrato e vedere
%se è mantenuto.
%
The Sharpe and the Calmar ratio are relevant also for portfolio
optimization, where they play the role of objective function. The
analysis of their relationship can be a basis for the study of
portfolio optimization and it will be studied elsewhere.

\section{The rising part of speculative bubbles}

Stock market indices are a particular portfolio,
based on a weighted mean of selected stocks. Buying/selling stock market indices has the meaning of buying/selling a previous selected financial product replica of
the index (Exchange Traded Funds, certificates).
The results of the previous sections can be used to examine the
behavior of stock market indices. The analysis of the performance
of the stock market indices can be more relevant than any other
portfolio because they can be selected in order to represent the
market risk and they can be used as a benchmark for
``beating the market''.
\newline
Our aim is to understand the behavior of MDD on raw data and to compare it with the results obtained on synthetic fBm. In particular we focus on stock market indices that have shown speculative bubbles due to endogenous causes, and we estimate the
risk measures on them.
\newline
Following Ausloos et al. \cite{abmv2,abmv}, and Sornette et al.
\cite{sornettecPhysRep,js,sorL}, we selected the indices that were
well modeled through the log-periodic correction to power law. In
all that cases the magnitude of the crash is proportional to the price, so it is
commonly suggested to use the logarithm of market index data
$Y(t)=log(p_t)$ \cite{js,sorL,js3,jsnasdaq}. We start performing
some statistical analysis and after we examine the behavior of
$E(MDD)$.

\subsection{Statistics on raw data}
Data selection is reported in Table \ref{tab:crashes}, together
with some statistical analysis: the mean of log-returns $\mu$
(i.e. the mean of the increments of $Y(t)$), their variance,
$\sigma$, the Hurst exponent $H$ measured on the log-returns
$\{Y(t+1)-Y(t)\}$ through the Detrended Fluctuation Analysis (DFA)
\cite{DFA1}, the Sharpe ratio and the Calmar ratio for the whole
time window considered. W.r.t. paper \cite{rotundonavarra} the
time series of Chile General, and the one corresponding to the
crash of Mexico Ipc in 1994 were not considered for the present
analysis because their length is lower than one year (256 data),
therefore they were not suitable for the same parameters to be
used for the DFA analysis performed on the other series. The time
series of Arg Burcap was retieved and added to the data set. The
Jarque-Bera test was performed on the data set listed in Table
\ref{tab:crashes} in order to test the Gaussianity of the
increments. The Gaussian hypothesis is rarely validated.
Therefore, the test was made on subwindows. For each $k$ and $j$
the Jarqe-Bera test was performed on the increments estimated on
$\{Y(t+1)-Y(t)\}_{t=k,k+j}$, i.e. on  time windows with starting
time $k$ and width $j$. The choice of the minimal time window
length is $256$ data, corresponding to the mean number of trading
days contained into a year. Figure \ref{fig:jball} reports the
starting time $k$ on the $x$-axis, the time window width $j$ on
the $y$-axis and it displays the result of the Jarque-Bera test
through a color palette.
 The upper right triangle of course is empty, and it has been drawn with a value equal to zero (white).
 In the lower triangle the white dots correspond to the acceptance
 of the Gaussian hypothesis, and the black ones to the rejection of the
 hypothesis.  Figure \ref{fig:jball} shows
that there are time windows in which the Gaussian hypothesis is
validated. It is not obvious that Gaussianity is reached for long
enough time windows. As an example in FTSE 100 there are time
windows of length 300 days that verify the Gaussianity test, while
longer time series reject this hypothesis, that is verified again
only for much longer time windows. In the same way,  the NASDAQ
100 (1998) shows Gaussian time windows of length 400-600.
A definitive convergence to the Gaussian case is shown for the
DAX, FTSE 100 (1987;1998), Hang Seng, Mexico Ipc, Nasdaq
(1987;2000).

\subsection{Estimating $MDD$ on raw data}
We aim at analyzing the situation of a  practioneer investing on a
real market portfolio, basing his risk measure on the sequence of
MDD actually happened rather than on an hypothetical $E(MDD)$.
We compare the MDD for the logarithm of the rising part of speculative bubbles for the indices considered in the present work with $E(MDD)$ of synthetic Brownian motions and $E(MDD)$ of synthetic fractional Brownian motions with Hurst exponent as given in Table \ref{tab:crashes} for each index.
The synthetic time series were simulated by using the same mean of returns and the same variance of the real data. For each index 1000 synthetic time series were generated for both type of process (Bm and fBm) and $E(MDD)$ was estimated averaging over the $MDD$ of each time series.
For both real data and synthetic time series $MDD$ and $E(MDD)$ were estimated by dividing each time window into six equally spaced intervals.
In Figure \ref{fig:figspecbubmdd} we show the comparison of the $E(MDD)$ of the synthetic Bm and fBm with the $MDD$ of real index data.
\newline
The difference of $MDD$ estimated on indices data from the
theoretical values can be addressed to different factors: 1.
$E(MDD)$ is a mean value, whilst the values reported for the
indices is a measure of the $MDD$ on a single trajectory. 2.
Deviations from the hypothesis of Gaussian increments (Jarque-Bera
test) can also give rise to deviation of $MDD$ from the
theoretical value of $E(MDD)$ for both Bm and fBm. 3. Another
factor of deviation from the theoretical values could be a result
of the error in the estimation of the Hurst exponent on short time
series \cite{weron}.
\newline
From the same figure it is possible to notice that the $MDD$ evaluated after a huge crash remains till an even bigger crash happens, leading to an overestimation of the risk on shorter sequences.

\section{Conclusions}
The present report deals with measures of performance considering
the maximum drawdown. Theoretical results on
$E(MDD(T))$, Sharpe and Calmar ratios available for Bm are
compared with empirical behavior of $E(MDD(T))$, Sharpe
and Calmar ratios on fBm. By extensive simulations we show the behavior of these performance measure as function of the Hurst exponent and the mean return sign. We show that, in the most relevant case of $\mu>0$, $H=0.5$ separates two different regime for $E(MDD(T))$ giving rise to a regime faster than logarithm when $H>0.5$ and slower than logarithm when $H<0.5$.
The results are then applied to the rising part
of speculative bubbles. Statistical analysis on such time series
is refined on their subsequences, and the distance of the
obtained results from the (f)Bm case is discussed.

\section*{Acknowledgements}
GR thanks Amir Atiya and Anna Maria D'Arcangelis
 for fruitful discussions on financial risk and related topics. FP
thanks Marcel Ausloos for pertinent comments and fruitful
discussions, and the European Commission Project
E2C2 FP6-2003-NEST-Path-012975  ``Extreme Events: Causes and Consequences'' for financial support.

\begin{table}[htbp]
\centering \caption{List of crashes. Data sets of rising part of
speculative bubbles are chosen from the rise of the bubble to the
expected crash time \cite{sorL}. W.r.t. paper
\cite{rotundonavarra} the time series of Chile General, and the
one corresponding to the crash of Mexico Ipc in 1994 were not
considered for the present analysis because their length is lower
than one year (256 data), therefore they were not suitable for the
same parameters to be used for the DFA analysis performed on the
other series. The time series of Arg Burcap was retieved and added
to the data set. In the table are reported respectively: index name, year of crash, mean of daily returns, standard deviation of daily returns, Sharpe ratio, Hurst exponent, maximum drawdown and Calmar ratio. All this quantities are estimated on the logarithm of each index.} \label{tab:crashes}
\begin{tabular}{|ll|l|l|l|l|l|l|}
\hline
Index & Year & $\mu$ & $\sigma$ & $Shrp$ & H & $MDD$ & $Calmar$
\\ \hline \hline
Arg Burcap & 1997 &  0.0012 & 0.0157 & 0.08 & 0.51 & 0.24 & 3.44
\\ \hline
Arg Merval & 1997   & 0.0015 & 0.0173 & 0.08 & 0.47 & 0.25 & 3.70
\\ \hline Brazil
Bovespa  &1997     & 0.0029 & 0.0136 & 0.21 & 0.65 & 0.12 & 8.46
\\ \hline
DAX 40 & 1998      & 0.0018 & 0.0124 & 0.14 & 0.63 & 0.20 & 4.43
 \\ \hline
FTSE 100 & 1987   & 0.0012 & 0.0082 & 0.15 & 0.60 & 0.12 & 5.64
\\ \hline
FTSE 100 & 1997   & 0.0008 & 0.0066 & 0.12 & 0.59 & 0.06 & 8.70
\\ \hline
FTSE 100 & 1998   & 0.0008 & 0.0076 & 0.11 & 0.61 & 0.12 & 5.71
 \\ \hline
Hang Seng &1994    & 0.0018 & 0.0133 & 0.14 & 0.55 & 0.26 & 4.31
\\ \hline
Hang Seng &1997 & 0.0010 & 0.0115 & 0.09 & 0.57 & 0.16 & 4.35
\\ \hline
Kuala Lumpur SE Emas &1994  & 0.0027 & 0.0095 & 0.28 & 0.75 & 0.10 & 10.06
\\ \hline
Mexico Ipc &1997   & 0.0017 & 0.0124 & 0.14 & 0.52 & 0.14 & 5.74
\\ \hline
Nasdaq 100 &1987  & 0.0012 & 0.0096 & 0.13 & 0.58 & 0.20 & 3.17 \\
\hline %per chiudere ciascuna riga
Nasdaq 100 &1998   & 0.0014 & 0.0149 & 0.09 & 0.56 & 0.20 & 6.40
 \\ \hline
Nasdaq 100 &2000 & 0.0023 & 0.0205 & 0.11 & 0.53 & 0.26 & 6.80
 \\ \hline
Venezuela SE Gen &1997 & 0.0041 & 0.0161 & 0.26 & 0.73 & 0.13 & 17.31
\\ \hline
\end{tabular}
\end{table}

\newpage

%\begin{figure}
%\includegraphics[height=10cm,width=12cm]{fig5.eps}
%\caption{ The plot shows the distance ($\mid
%E(MDD(T))_{H}-E(MDD(T))_{0.5}\mid$) of each curve from the one
%drawn for $H=0.5$.} \label{fig5}
%\end{figure}

\newpage
\begin{figure}
\includegraphics[height=15cm,width=12cm]{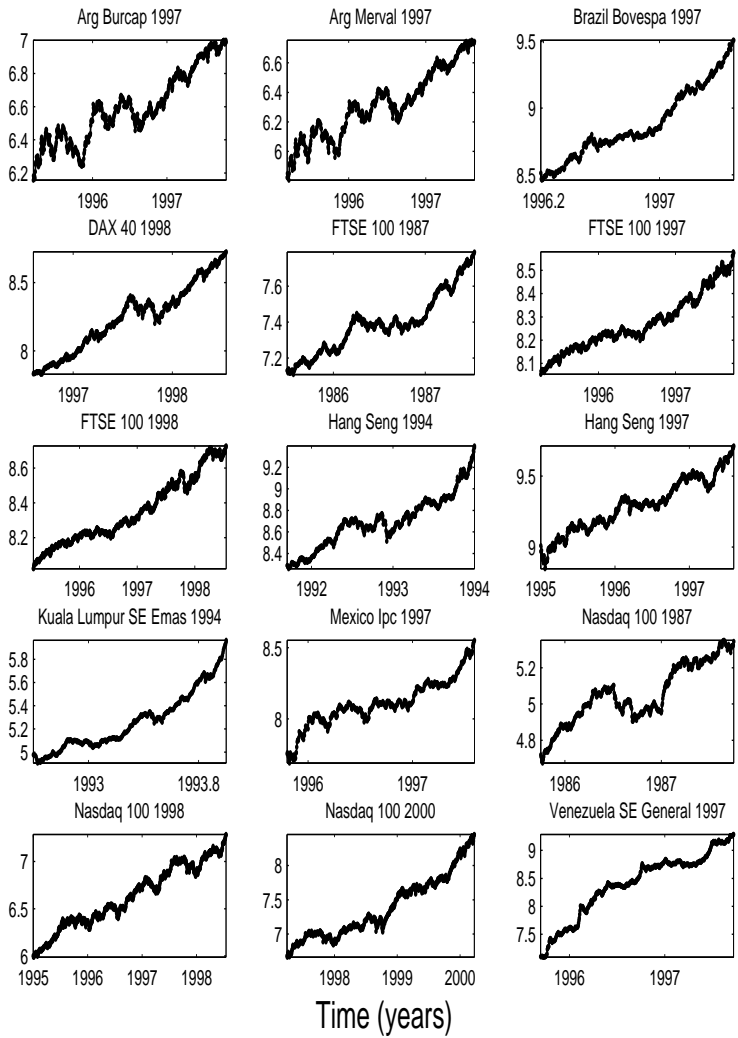}

%\centerline{\epsffile{fig_spec_bub.eps}}

\caption{Rising part of speculative bubbles for the indices
considered in the present work, and listed in Table
\ref{tab:crashes}.}\label{bubbles}
\end{figure}

\newpage
\begin{figure}
\includegraphics[height=10cm,width=12cm]{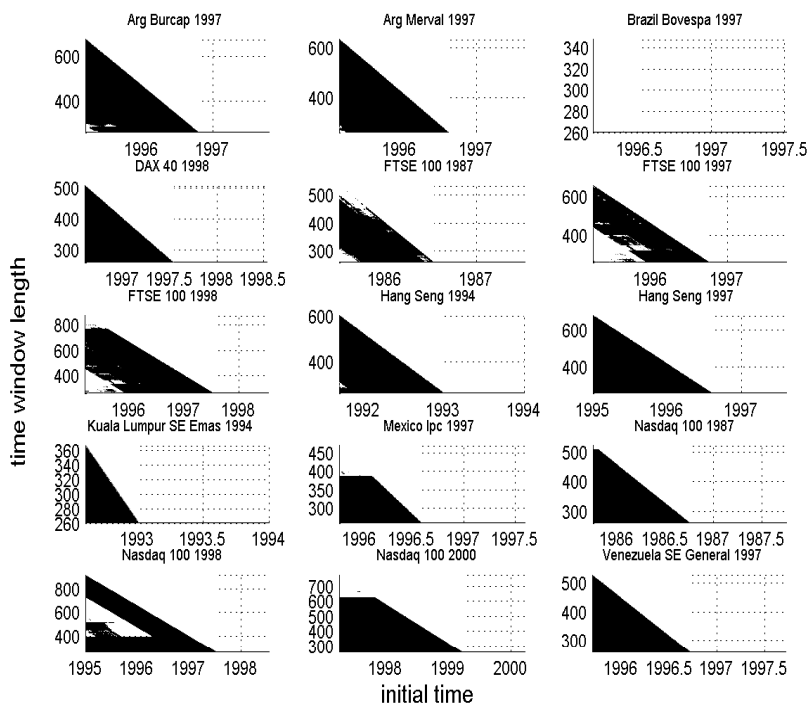}
\caption{Jarque-Bera test on the rising part of speculative
bubbles. The x-axis report the initial day of the time window; the
y-axis reports the time length ($>$256 days). The upper triangle
does not contain information because of the finite size of the
time series. The lower triangle reports the result of the
Jarque-Bera test. For each $(x,y)$ white corresponds to
the Gaussian case; black rejects the Gaussian hypothesis on
the time window $(x,x+y)$.} \label{fig:jball}
\end{figure}

 \begin{figure}
\includegraphics[height=10cm,width=12cm]{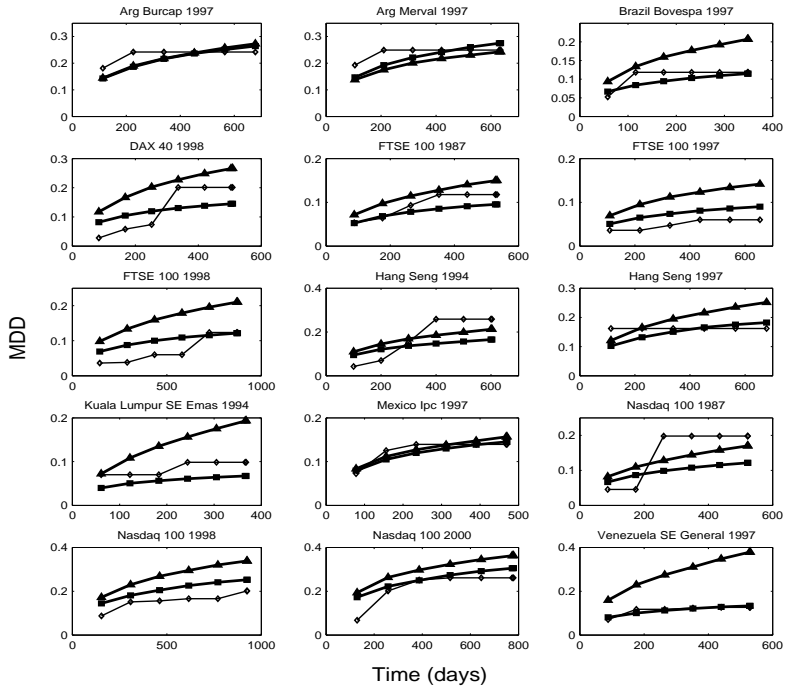}
%\centerline{\epsffile{fig_spec_bub_mdd.eps}}
\caption{$MDD$ for the logarithm of the rising part of speculative
bubbles for the indices considered in the present work. MDD
(diamonds) are compared with $E(MDD)$ of synthetic Brownian
motions (squares) and $E(MDD)$ of synthetic fractional Brownian
motions (triangles) with Hurst exponent as given in Table
\ref{tab:crashes} for each index. $E(MDD)$ for Bm and fBm is
averaged over 1000 simulation. $x$-axes report the time expressed
in days in a log scale.}\label{fig:figspecbubmdd}
\end{figure}

\end{document}